%% file: 2014_SPIE_ALICE.tex
\documentclass[]{spie}  
\usepackage[]{graphicx}
\usepackage[]{times}
\usepackage[utf8]{inputenc}
\usepackage[T1]{fontenc}
\usepackage{lmodern}
\usepackage{gensymb}        
\usepackage[usenames,dvipsnames]{color}

\newcommand{\todo}[1]{\textbf{\textcolor{red}{#1}}}
\newcommand{\notSPIE}[1]{}
\newcommand{\notspie}[1]{}

\newcommand{\larry}[1]{\textbf{\textcolor{Melon}{#1}}}
\newcommand{\elo}[1]{\textbf{\textcolor{Turquoise}{#1}}}

\title{Archival Legacy Investigations of Circumstellar Environments: Overview and First Results}

\author{
    Élodie Choquet\supit{a}, 
    Laurent {Pueyo}\supit{a}, 
    J. Brendan {Hagan}\supit{a,b}, 
    Elena Gofas-Salas\supit{a,f}, 
    Abhijith {Rajan}\supit{g},
    Christine {Chen}\supit{a}, 
    Marshall D. {Perrin}\supit{a}, 
    John {Debes}\supit{a}, 
    David {Golimowski}\supit{a}, 
    Dean C. {Hines}\supit{a}, 
    Mamadou {N'Diaye}\supit{a}, 
    Glenn {Schneider}\supit{c}, 
    Dimitri {Mawet}\supit{e}, 
    Christian {Marois}\supit{d}, 
    Rémi {Soummer}\supit{a, \dag}
\skiplinehalf
\supit{a}Space Telescope Science Institute, 3700 San Martin Drive, Baltimore MD 21218, USA; \\
\supit{b}Purdue University, West Lafayette, IN 47907, USA; \\
\supit{c}Steward Observatory, University of Arizona, North Cherry Avenue, Tucson AZ 85721, USA; \\
\supit{d}NRC Herzberg Institute of Astrophysics, West Saanich Road, Victoria, BC V9E 2E7, Canada; \\
\supit{e}ESO, Alonso de Cordova 3107, Casilla 19001, Santiago 19, Chile; \\
\supit{f}Institut d'Optique Graduate School, Palaiseau Saint-\'Etienne and Bordeaux, France; \\
\supit{g}Arizona State University, Phoenix, AZ 85004, USA; \\
}

\authorinfo{Correspondence should be sent to choquet@stsci.edu, \dag: Principal Investigator}

\begin{document} 
  \maketitle 

\begin{abstract}
We are currently conducting a comprehensive and consistent re-processing of archival HST-NICMOS coronagraphic surveys using advanced PSF subtraction methods, entitled the Archival Legacy Investigations of Circumstellar Environments program (ALICE, HST/AR 12652). This virtual campaign of about 400 targets has already produced numerous new detections of previously unidentified point sources and circumstellar structures. We present five newly spatially resolved debris disks revealed in scattered light by our analysis of the archival data. These images provide new views of material around young solar-type stars at ages corresponding to the period of terrestrial planet formation in our solar system.  We have also detected several new candidate substellar companions, for which there are ongoing followup campaigns (HST/WFC3 and VLT/SINFONI in ADI mode). Since the methods developed as part of ALICE are directly applicable to future missions (JWST, AFTA coronagraph) we emphasize the importance of devising optimal PSF subtraction methods for upcoming coronagraphic imaging missions. We describe efforts in defining direct imaging high-level science products (HLSP) standards that can be applicable to other coronagraphic campaigns, including ground-based (e.g., Gemini Planet Imager), and future space instruments (e.g., JWST).  ALICE will deliver a first release of HLSPs to the community through the MAST archive at STScI in 2014.
\end{abstract}

\\
\notSPIE{Note: blue parts will be removed from SPIE (but maybe kept for another paper and/or ALICE manual).\\}

\section{INTRODUCTION} \label{sec:intro} 

Since the detection of the first exoplanet\cite{Mayor1995}, our understanding of planetary systems and their formation has been revolutionized by the discovery of more than a thousand other extra-solar planets. Radial velocity and transit measurements have been the most successful methods, with hundreds of discoveries through several surveys over the past 15 years\cite{CollierCameron2007,Lovis2011,Batalha2013}. 
However, these methods are limited to the detection of planets relatively close to their host stars, and need very long time-scales to confirm planets on very wide orbits. Complementary, direct imaging provides observations of the whole environment of nearby stars at a glance, enabling detection and characterization of dusty disk and of systems with massive planets both spatially and spectraly\cite{Soummer2011,Konopacky2013,Oppenheimer2013,Hinkley2013}.

However, such observations are very challenging due to the large contrast between the circumstellar objects and their host star, when closer than 1''. Despite multiple surveys conducted with telescopes equipped with coronagraphic systems, only a handful of bright debris disk have been imaged until the late 2000's \cite{Golimowski2011}, and the first images of exoplanets were obtained only six years ago\cite{Marois2008,Kalas2008,Lagrange2010}. These discoveries were only possible thanks to careful star point spread function (PSF) subtraction during post-processing. At this time, the most common PSF subtraction method merely consisted in subtracting the image of a reference star, or a median image of a collection of reference stars from the target images. However, this method is still  not optimal to reveal faint structures less than 2'' from the host star, due to slight variations of the wavefront (e.g. thermal breathing for space telescopes, residual errors form atmospheric turbulence for ground based telescopes with adaptive optics). For instance, in the case of the Hubble Space Telescope (HST), well-known for its stability \cite{Krist1998}, small temporal variations\cite{2006SPIE.6270E..54L} (focus breathing and cold mask shifts induce by periodic thermal variations along an HST orbit) affect classical PSF subtraction methods down to a level that prevents coronagraphic imaging to reach the fundamental limit of the speckle's photon noise within the inner $\sim$ 2'' of the PSF.

New post-processing methods based on the use of large libraries of reference stars have opened a new era in high-contrast exploration of circumstellar environments. The reference library is here used as a sample of optical path random realizations to generate a synthetic PSF as close to science image star PSF as possible. 
Since the development of the \emph{Locally Optimized Combination of Images} (LOCI) algorithm \cite{Lafreniere2007}, and more recently with the use of Principal Component Analysis algorithms \cite{Soummer2012,Amara2012}, about 40 substellar companions have been imaged around nearby stars\footnote{Source: The Extrasolar Planets Encyclopedia, http://exoplanet.eu}.

These post-processing methods have now became standard techniques for high contrast imaging. However, numerous coronagraphic programs have been carried out before 2010 without such optimized processing, and thus these programs potentially have unseen circumstellar structures that are within reach of the contrast achieved with these recent techniques. It has actually been demonstrated with the re-discovery of the HR~8799 planetary system\cite{Lafreniere2009,Soummer2011} using HST NICMOS archival data acquired 10 years before their first images\cite{Marois2008}.

In this context, we initiated a consistent and comprehensive re-processing of the entire NICMOS coronagraphic archive using advanced post-processing techniques, in order to reveal hidden objects and reach better detection limits with state-of-the-art algorithms. With about 400 targets in the archive, large reference libraries can be easily assembled to optimize the computation of the most accurate synthetic PSF for each science image.  
In this paper, we present an overview of this project, entitled \emph{Archival Legacy Investigation of Circumstellar Environment}\footnote{HST-AR-12652, PI: R. Soummer} (ALICE). We describe the NICMOS archive content in Sec.~\ref{sec:archive}, then we detail the architecture of the ALICE pipeline in Sec.~\ref{sec:pipeline}. First results of the project are presented in Sec.~\ref{sec:results}. We then conclude by presenting the prospects of the ALICE project in Sec.~\ref{sec:conc}.

\section{NICMOS ARCHIVE CONTENT}\label{sec:archive} 

Our database contains all the NICMOS images which have been reprocessed through the HST archival LAPLACE program\footnote{HST-AR-11279, PI: G. Schneider} (\emph{Legacy Archive PSF Library And Circumstellar Environments})\cite{Schneider2010}, i.e. all NICMOS coronagraphic images (acquired with the NIC2-CORON aperture) from cycle 7 through cycle 15, and the additional program 11155 (cycle 16, PI: M. Perrin). The LAPLACE reprocessing consists of an improved recalibration of the NICMOS coronagraphic data using contemporary flat field frames, observed dark frames and a better bad-pixel correction. This recalibration is crucial to build optimal reference libraries needed for advanced PSF subtraction algorithms as will be described in Sec.~\ref{sec:pipeline}. The additional cycle 16 program has been added to our database to analyze the interface of the ALICE pipeline with non-LAPLACE data. Although included in the following description of our database, none of the results presented in this paper are obtained with data from this program.

\begin{figure}
   \begin{center}
   \begin{tabular}{c}
   \includegraphics[height=5cm]{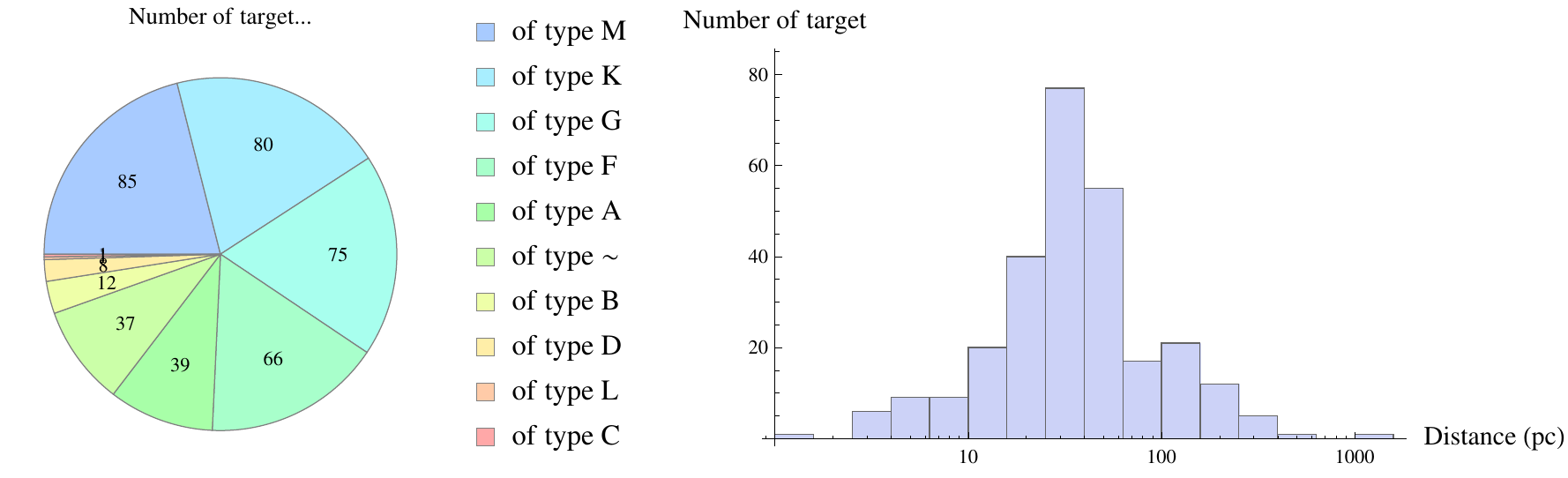}
   \end{tabular}
   \end{center}
   \caption{\textbf{Left}: Distribution of spectral types of our target list. \textbf{Right}: Histogram of the known distances of the stars in our database. 131 targets have unreferenced parallax measurements, out of the 405 stars in our target-list. 
Source: the SIMBAD database, http://simbad.u-strasbg.fr/simbad. \label{fig:specTypeAndDistance}}
   \end{figure}

The database contains images from 405 stars. Their spectral type distribution is roughly equally balanced between types M, K, G, and other spectral types, for stars with identified spectral type\footnote{Source: The Simbad astronomical database, http://simbad.u-strasbg.fr/simbad/} (see Fig.~\ref{fig:specTypeAndDistance}, left). About 50~\% of the stars with known parallax\footnote{Source: The Simbad astronomical database, http://simbad.u-strasbg.fr/simbad/} in our target list are within 35~pc from Earth (see Fig.~\ref{fig:specTypeAndDistance}, right).

\begin{figure}
   \begin{center}
   \begin{tabular}{c}
   \includegraphics[height=7cm]{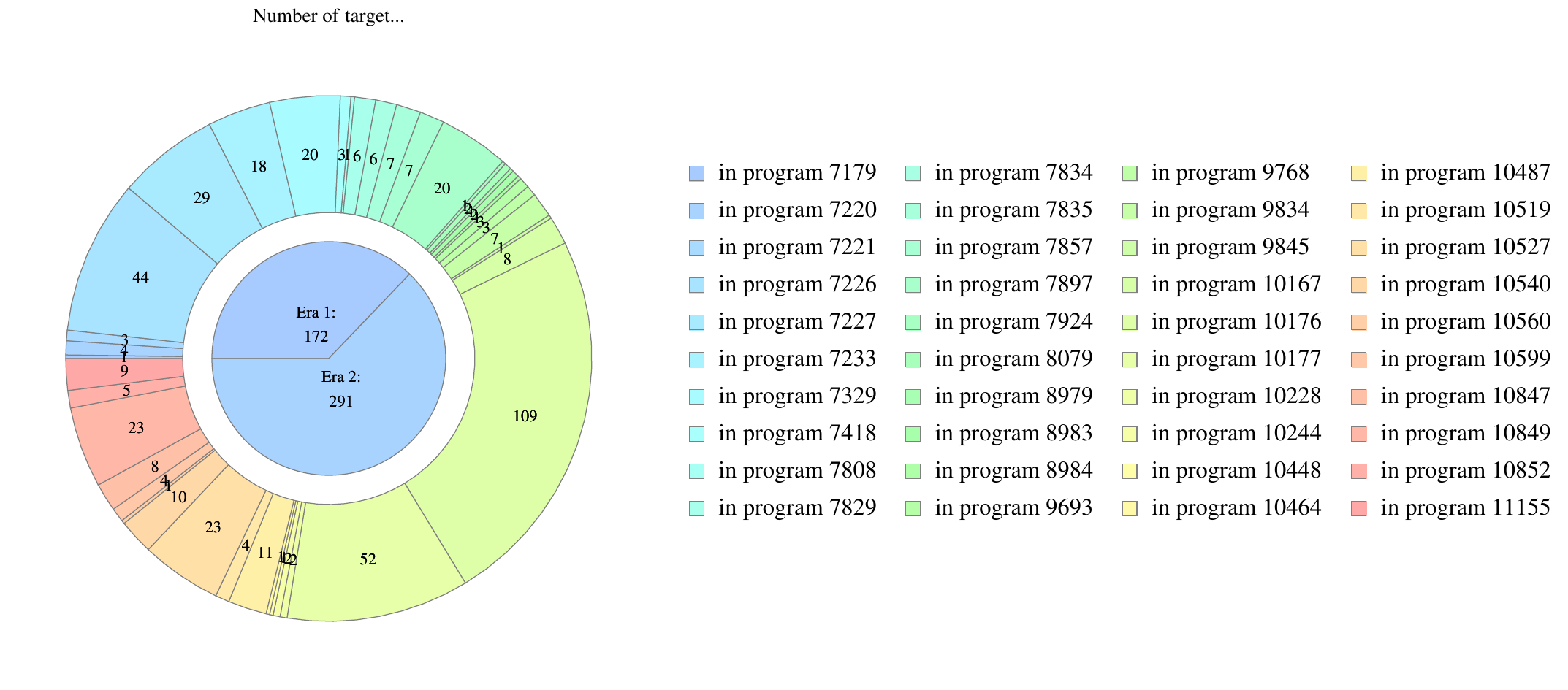}
   \end{tabular}
   \end{center}
   \caption{Number of target per NICMOS coronagraphic program. The programs from 7179 to 8079 were conducted during the first Era of NICMOS. The programs from 8979 to 11155 were conducted between 2002 and 2008, after replacement of the nitrogen-cooler by a  cryogenic neon cryocooler. All these data have been reprocessed as part of the LAPLACE program, except program 11155 which have been recently added to our database to investigate on the compatibility between the ALICE pipeline and non-LAPLACE NICMOS data. \label{fig:targetParProgram}}
   \end{figure}

The images in our database are distributed into 40 programs (See Fig.~\ref{fig:targetParProgram}). 16 of them have been carried out between 1997 and 1999, during the first era of NICMOS during which the instrument was cooled with a nitrogen-ice cryogenic dewar. The cooling system was replaced by a cryogenic neon cryocooler in 2002 during Hubble Service Mission 3B, after the dewar ran out of nitrogen in 1999. The last 24 programs of our database were conducted during this second era of the instrument, which represents 75~\% of our database in number of images. \notspie{23 targets have been observed in both instrument eras.} This distinction between both eras is an important parameter in the image selection to assemble the reference libraries, as will be explained in Sec.~\ref{sec:pipeline}.

Most of the images in our database are part of a survey program conducted to discover planets, debris disks, or protoplanetary disks. A brief description of these programs is provided in Table~\ref{tab:surveys}. Among the 405 stars in our database, 48 have been observed in several programs (12~\% of our target list). If a companion candidate is detected around one of them, whether or not is it gravitationally bound to the parent star can potentially be confirmed using NICMOS data from another program, provided that the temporal baseline between both datasets and the star proper motion are important enough to differ the companion motion from a background object.

\begin{table}
\caption{NICMOS coronagraphic surveys conducted to search for planets and disks. Only surveys with more than 150 images are described.}
\label{tab:surveys}
\begin{center}       
\begin{tabular}{lllll} 
Program 	&	PI		& \# Target		&NICMOS Filters & Type of survey \\	
\hline
7226	& E. Becklin	&44			&F110W, F160W, F165M, F180M	&Planets (young and nearby)\\   				
7227	& G. Schneider &29			&F160W 						&Planets (around M stars)\\			
7233	&B. Smith		&18			&F110W, F160W, F204M, F237M	&Disks (MS stars, IRAS excess)\\
7834	&R. Rebolo	&6			&F110W, F160W, F180M, F207M	&Planets (young, nearby, late)\\	
10167	&A. Weinberger&8			&F171M, F180M, F204M, F222M	&Disks (known disks)\\				
10176	&I. Song		&109		&F160W						&Planets (young nearby)\\			
10177	&G. Schneider	&52			&F110W, F160W				&Disks (IRAS excess)\\	
		&D. Ardila	&11			&F110W						&Disks (Beta pic moving group)\\		
10519	&J. Simpson	&4			&POL0L, POL120L, POL240L		&Disks (YSO, massive)\\				
10527	&D. Hines		&23			&F110W						&Disks (Spitzer excess)\\			
10540	&A. Weinberger&10			&F110W 						&Disks (IR excess)\\				
10847	&D. Hines		&8			&POL0L, POL120L, POL240L		&Disks (known disks)\\				
10849	&S. Metchev	&23			&F110W						&Disks (Spitzer excess)\\			
11155	&M. Perrin	&9			&F110W, POL0L, POL120L, POL240L&Disks (Herbig Ae)\\				

\end{tabular}
\end{center}
\end{table}

The vast majority of the images in our sample have been acquired with the J and H wide-band filter (67~\% of the target list in the F160W filter, and 40~\% in the F110W filter). The other filters were marginally used with the coronagraphic mode of NICMOS\notSPIE{ (See Fig.~\ref{fig:NombreTargetParFilter})}. Yet, 95 targets (23~\% of the target list) have been observed at least two different filters. The comprehensive reprocessing of these targets will thus provide color measurements of any object detected around the host star (co-moving companion, disk, as well as background object). \notspie{In particular a handful of targets have been observed with more than 5 different filters, whose reprocessing will result in low-resolution spectral density distributions (SED).}

\notSPIE{\begin{figure}
   \begin{center}
   \begin{tabular}{c}
   \includegraphics[height=7cm]{FigNombreTargetParFilter.pdf}
   \end{tabular}
   \end{center}
   \caption{\notSPIE{Number of target per NICMOS filter.\label{fig:NombreTargetParFilter}}}
   \end{figure} }

Although the majority of our sample has been observed with several different orientations of the telescope (82~\% of them in exactly two orients of HST), 82 stars have been only observed for a single orientation of the telescope, which significantly increase the false-alarm probability of faint point source detection. \notSPIE{Without  additional observations at different telescope orientation, it is not possible to distinguish between a $1~\sigma$ point source and a residual speckle. \larry{I would remove this, unless we can really quantify what we are talking about.} \elo{I think it is still a safe statement here, because by definition a 1$\sigma$ point source can be anything, right?}}

\section{THE ALICE PIPELINE}\label{sec:pipeline}

The ALICE pipeline has been designed to enable processing by several users of any coding level. It is entirely developed in \emph{Wolfram Language} with \emph{Mathematica} and utilizes several graphical user interfaces (GUIs) to ease its exploitation. Several tools have been developed to keep tracks of the user activities and of the intermediate files generated.

The pipeline is structured into three main sections:
\begin{itemize}
\item \textbf{Data preparation}: In this part, images from several programs are gathered, aligned, and sorted to create reference image cubes that are necessary for advanced PSF subtraction algorithms.
\item \textbf{Data reduction}: The targets to be reduced are selected, a parameter space is defined to optimize the reduction, then the selected images are reduced using the KLIP algorithm \cite{Soummer2012}.
\item \textbf{Data analysis}: All reduced images of a target are examined and combined, and the reduction parameter space is explored to determine the optimal reduced image. Residual speckles are analyzed to determine if they are real point sources or star residuals, then final science products are generated.
\end{itemize}

Several log files keep track of the operations performed on the data. The main log is a global workflow file which gives an overall view of the processing status of each program--target--filter combination. A log entry is added each time a target starts or ends being processed by one of the main section of the pipeline \notspie{("Prepping", "Prepped", "Reducing", "Reduced, "Analyzing", "Completed")}. Additional log entries are also generated at intermediate steps of the process. For each entry, the workflow indicates the user, pipeline version, and name of files generated or used at the different steps. The workflow is thus a key functionality for managing such a large program, including multi-user activities, or following the status of particular targets through each processing steps until final product delivery. \notSPIE{A sample of the workflow file is presented in Table.~\ref{tab:workflow}}

\notSPIE{\begin{table}
\caption{\notSPIE{Sampled of the workflow file used to keep track of the processing status of every target in the ALICE pipeline. The field providing the product filenames has been omitted here for sake of clarity.}}
\label{tab:workflow}
\begin{center}       
\begin{tabular}{llllllll} 
Program& Target&Filter&Status		&Workflow Step		&User	
&Date&Pipeline\\
\hline
10176&HIP3556&F160W&Not Prepped	&&&&\\
10176&HIP3556&F160W&Prepping		&&&&\\
10176&HIP3556&F160W&			&Bad images flagged	&Elodie
&2014-03-04&ALICE 2.0.8\\
10176&HIP3556&F160W&			&Aligned				&Elodie
&2014-03-04&ALICE 2.0.8\\
10176&HIP3556&F160W&			&Bad aligns flagged		&Elodie
&2014-03-05&ALICE 2.0.8\\
10176&HIP3556&F160W&			&Global LOCI			&Elodie
&2014-03-05&ALICE 2.0.8\\
10176&HIP3556&F160W&			&Bad Ref Flagged		&Elodie
&2014-03-05&ALICE 2.0.8\\
10176&HIP3556&F160W&Prepped		&&&&\\
10176&HIP3556&F160W&Reducing		&&&&\\
10176&HIP3556&F160W&			&RunOptimSearchParam&Elodie
&2014-04-09&ALICE 2.0.8\\
10176&HIP3556&F160W&Reduced		&&&&\\
10176&HIP3556&F160W&Analyzing	&&&&\\
10176&HIP3556&F160W&			&Combined products&Elodie
&2014-05-01&ALICE 2.0.8\\
10176&HIP3556&F160W&Complete		&&&&\\
\end{tabular}
\end{center}
\end{table} }

\subsection{Construction of the Reference Libraries}

The first part of the pipeline consists in building reference cubes that will be used by advanced PSF subtraction to reduce the science images. This is the most critical part of the pipeline, for a mis-prepared reference cube will result in a poor PSF subtraction at the next step of the pipeline. This part is also the most computation-time consuming.

\subsubsection{Image selection}
The first step to select potential reference images by combing several archival programs, including the one containing the science target to be reduced. Ideally, one wants a large selection of PSFs similar to the target image to achieve an optimal subtraction, typically larger than 200 images. 
To achieve efficient PSF post-processing, two requirements have to be met:
\begin{itemize}
\item The selected programs have to have been conducted during identical NICMOS cooling eras, so either before or after 2002, but not a mix.  Since the cooling system have been modified between the two eras, the detector responses (dark frames, bad pixels, flat fields) are not fully comparable despite similar reprocessing by the LAPLACE program \cite{Schneider2010}. References from one cooling era do not provide good matches to reduced a science image from the other instrument era. 
\item All the images in the reference cube have to be acquired with a unique and identical spectral filter. The instrument PSF is indeed wavelength-dependent, as well as the radial position of star-induced residual speckles. PSF subtraction from reference images with different filter systematically results in poor starlight suppression.
\end{itemize}

The selected images are then cropped to a squared field of view, hereafter the ALICE field of view (typically 6'' across). The field of view is centered on the star position, which is identified using the location predicated from the LAPLACE processing (FITS file keywords "TARSIAFX" and "TARSIAFY"). \notspie{A precision of 760~$\mu$as is predicted by [\citenum{Schneider2010}] for these values, estimated from non-coronagraphic target acquisition centroid position and telescope slew telemetry after target re-centering on the coronagraphic mask, and thus assuming that the instrument is stable to sub-milliarcsecond (mas) astrometric precision during the procedure.}

\begin{figure}
   \begin{center}
   \begin{tabular}{c}
   \includegraphics[height=4.2cm]{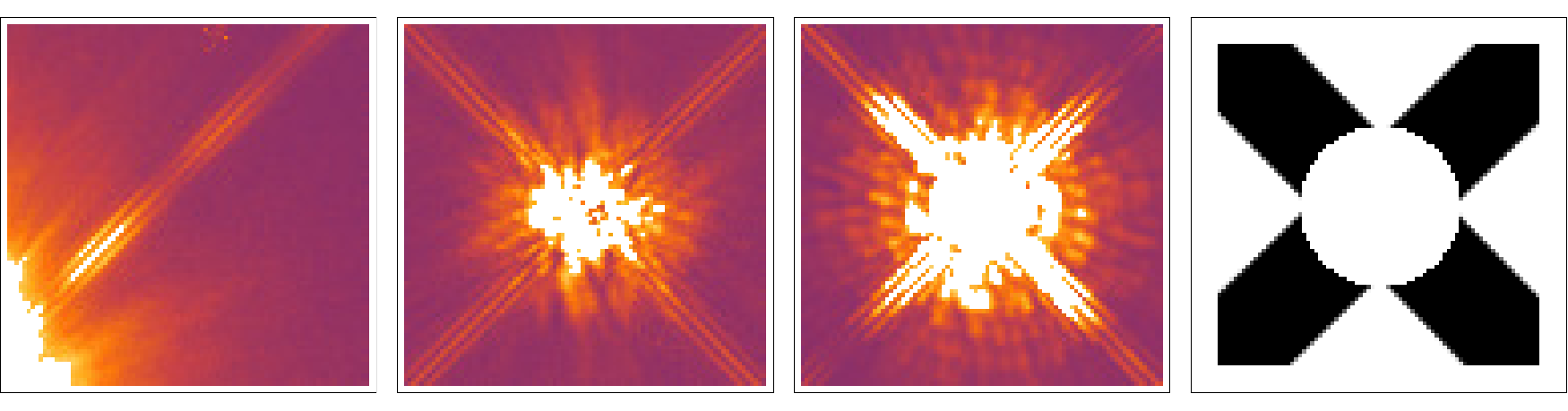}
   \end{tabular}
   \end{center}
   \caption{\textbf{Left}: typical bad acquisition image dismissed from the reference cube. \textbf{Mid-left}: potential good reference image. \textbf{mid-right}: F160W TinyTIM synthetic PSF, used as the initial reference for alignment. \textbf{Right}: Typical mask used for the alignment of the reference cube.\label{fig:rawImages}}
   \end{figure}

Finally, the images are examined to flag and discard the obvious failed acquisitions. About 5~\% of all images in our database has been flagged as a bad image. Fig.~\ref{fig:rawImages} presents examples of a bad acquisition image (left) and of a potentially good reference image (mid-left). Note that no further correction of bad pixel is applied to the images. Only a few bad pixels are occasionally present in the images thanks to the LAPLACE reprocessing, which do not significantly alter the performance of the data processing.

\subsubsection{Image alignment}
The next step of a reference cube preparation is the fine alignment of all its images. This process is the most critical to achieve efficient PSF subtraction. The field centering on the star location predicted by the LAPLACE reprocessing is indeed not precise enough, because of instrumental variations during the spacecraft slew between the non-coronagraphic target acquisitions used to find the star centroid and the coronagraphic acquisitions.
\notSPIE{\todo{[Do a plot showing the star position predicted by LAPLACE and the one computed with ALICE.]}\elo{That would be a really interesting plot. But maybe to much details for this spie paper.}
}

The alignment is performed using a synthetic Tiny Tim PSF \cite{Krist2011} simulated in the F160W filter, whose known position is at the center of the ALICE field of view. We define this as the absolute reference position for the alignment (see Fig.~\ref{fig:rawImages}, mid-right). However, the model PSF is simulated without coronagraphic mask and is thus quite different from the cube images. We thus use a mask blocking out the core of the PSF and align the images only using the diffraction pattern of the telescope struts, which are clearly visible in NICMOS coronagraphic images (see Fig.~\ref{fig:rawImages}, right, for a typical alignment mask). Since the unobscured Tiny Tim  PSF is still slightly different from real images, we perform the alignment in two steps:
\begin{enumerate}
\item A median image from the reference cube is aligned with the Tiny Tim PSF;
\item All the images in the cube are then aligned with this aligned median image as reference.
\end{enumerate}
The alignment mask is used in both steps for every image, to focus the alignment on the diffracted spiders and avoid mis-alignments in case of bright close binary or disk contamination. In addition, the images can be convolved by a gaussian filter (standard deviation chosen by the user), to account for the fact that NICMOS images are Nyquist-sampled at 1.7~$\mu$m, and consequently undersampled at shorter wavelengths.

The default alignment algorithm used in the pipeline finds the position of the field of view maximizing the cross-correlation between an image and the reference image for alignment, filtered by the mask. 
\notspie{ This is done by minimizing the cost function:
\begin{equation}
f(x, y) = \log\left(\sum^{N_{pix}} \left(M.I(x,y)-M.R\right)^2\right)
\end{equation}
over the horizontal and vertical shifts $(x,y)$, with $I$ the image, $R$ the reference for alignment, and $M$ the mask.} 
All aligned images are then individually examined by the user to check the alignment with the reference image. Normalization of all the image of the cube is also performed to ease the examination of the alignment by the user. 
\notspie{This is done by finding the root of the derivative of blabla (no longer in the findminimum.)}
\notspie{The final value $f(x,y)$ obtained for the aligned images are also provided to help the alignment examination.}

Because of the third-order interpolation used for alignment, this process the most time-consuming of the pipeline. This method is particularly efficient for images in the F110W and F160W filter, where the diffraction pattern of the spiders is clearly visible. At longer wavelengths, the intensity of quasi-static speckles becomes dominant and impairs the alignment quality. Bad pixels can also occasionally impair the alignment when they spread over several pixels during the procedure because of sub-pixel interpolation. For these reasons, a careful examination of the alignment of all images is necessary at this step. We are currently analyzing different alignment algorithms to offer alternative solutions to the users of the ALICE pipeline depending on their need of celerity or robustness\cite{Gratadour2005}.

\subsubsection{Detection of bad references}

The final step in the preparation of a reference cube is the identification of bad reference targets. To properly subtract a synthetic PSF  with advanced algorithms, the PSF libraries must indeed be clean of any science components. Images in the reference cube with disks or point sources can indeed leave inconvenient artifacts in the final reduced images.

\begin{figure}[p]
   \begin{center}
   \begin{tabular}{c}
   \includegraphics[width=0.9\linewidth]{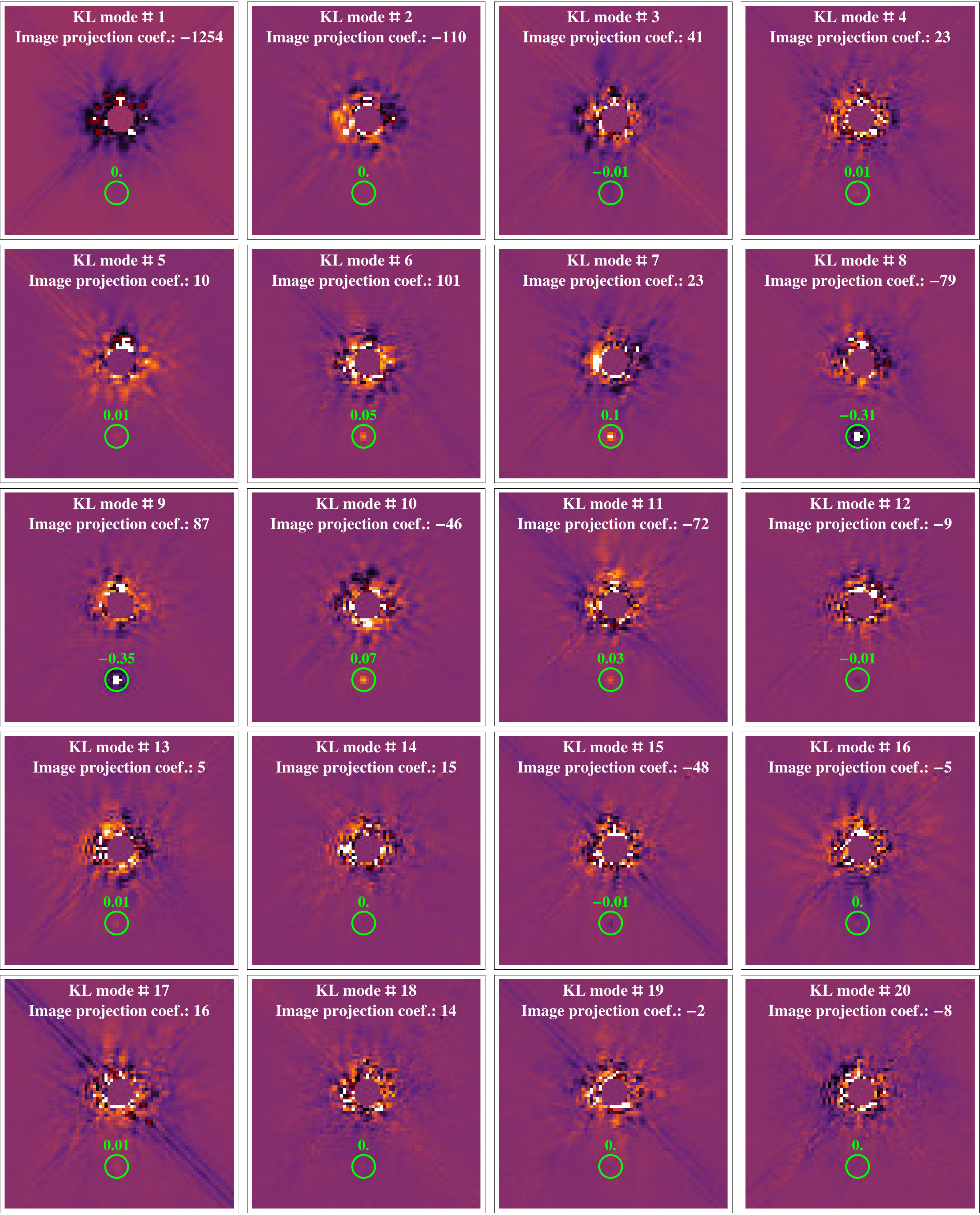}
   \end{tabular}
   \end{center}
   \caption{First twenty Karhunen-Loève modes of a toy reference cube. 
   The cube contains the known bad reference TWA 5, which is a binary star\cite{Lowrance2005}. The compagnon of the bad reference appears both with positive and negative values in the first KL modes of the cube, at positions indicated by green circles. The value above the circles are the pixel values of the modes at the center of the circles. The coefficients of the projection of a toy science image (known as being a good reference PSF) on this KL basis are displayed at the top of each image.\label{fig:KLModes}}
   \end{figure}

\begin{figure}
   \begin{center}
   \begin{tabular}{c}
   \includegraphics[width=0.6\linewidth]{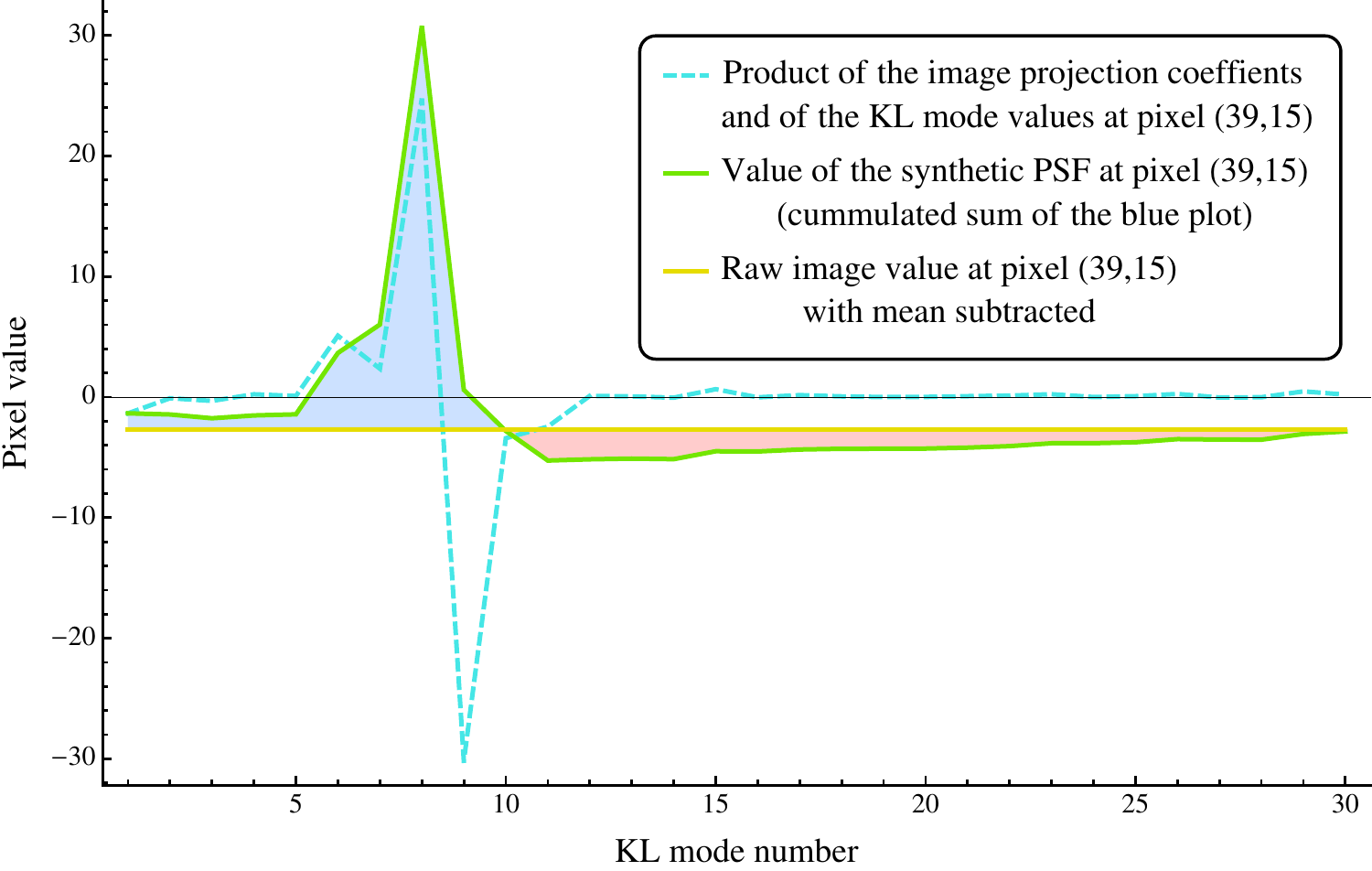}
   \end{tabular}
   \end{center}
   \caption{Details on the reduction of a toy science image with the KLIP algorithm\cite{Soummer2012}, using a reference cube with a bad target. The bad target is a binary star where the companion is a pixel position (39,15). The dashed blue curve shows the product between the KL mode value for this pixel (green values in Fig.~\ref{fig:KLModes}) and the coefficient of the projection of the science image on the modes (white values in Fig.~\ref{fig:KLModes}). The green curve is the cumulated sum of the later plot, and gives the value of the synthetic PSF provided by KLIP with increasing truncated mode value. The yellow curve shows the value of this pixel in the raw science image with mean subtracted. The difference between the two latter curves show that the use of this staled reference cube leads to the presence of a negative artifact (blue filling) for truncated values lower than 10 and to a positive artifact (red filling) for truncated values between 11 and 30, at the pixel position of the bad reference companion.\label{fig:subrtractionKLIP}}
   \end{figure}

To detect bad references in a PSF library, we first reduce each of its images using the cube itself as the PSF library (removing all other images of the target being reduced to avoid self-subtraction). This helps reveals if companions or disks are hidden by the bright PSF. At this step, we use the LOCI algorithm \cite{Lafreniere2007} on a single zone, the full ALICE field of view (masking only the center of the image with a small mask of typical radius of $\sim 5$ pixels), with a standard matrix regularization, identical for all the reduced images in the cube. For this first processing LOCI is preferable to PCA algorithms, because artifacts due to bad references always appear in negative in the reduced images, due to very aggressive subtraction of the LOCI algorithm. With PCA, the presence of a companion in a reference image appears in several Karhunen-Loève (KL) components of different ranks, with both negative and positive values (see Fig.~\ref{fig:KLModes}). The projection of the science image on the KL basis also independently provides negative and positive coefficients, which can result in either negative or positive artifacts in the PSF subtraction, depending of the number of KL modes used to compute the synthetic PSF (See Fig.~\ref{fig:subrtractionKLIP}). Fig.~\ref{fig:subrtractionKLIP} presents examples of PSF subtraction with the two algorithms, and shows that using PCA can be misleading to identify bad references, with positive artifacts that looks like faint companions.

The typical time to reduce one full image using LOCI with matrix regularization is $\sim 180$~ms on a 2.8~GHz laptop. When all the images in the cube have been reduced, they are carefully examined to identify targets with circumstellar material, which are then flagged as bad references. The process (reduction, examination, identification of bad references) is repeated several time by using the reference cube cleaned of the bad references, to refine the reduction and detect fainter companions. Note that some of these bad references are also interesting science target, whose reduction will be refined and optimized in the next section of the pipeline. 

\begin{figure}
   \begin{center}
   \begin{tabular}{c}
   \includegraphics[width=0.9\linewidth]{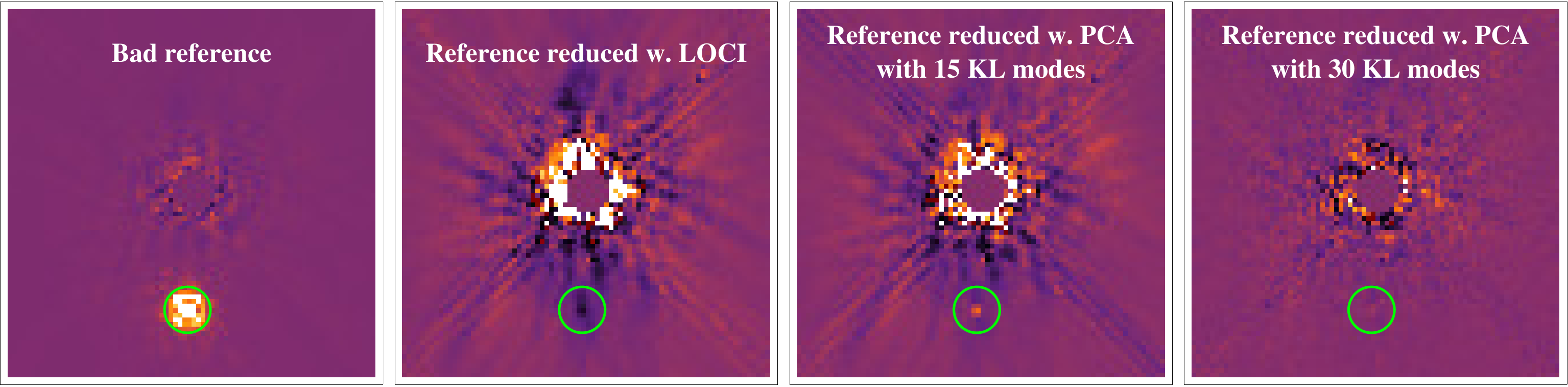}
   \end{tabular}
   \end{center}
   \caption{Impact of different reductions with a staled reference cube. \textbf{Left}: image of the bad reference in the PSF library. \textbf{Mid-Left}: science image reduced with LOCI, showing the presence of a negative artifact due to the bad reference in the PSF cube. \textbf{Mid-right}: science image reduced with the KLIP algorithm with 15 KL modes used to compute the synthetic PSF; the presence of the bad reference in the library adds a positive artifact which looks like a faint companion. \textbf{Right}: reduction with KLIP, with 30 KL modes to build the synthetic PSF; no artifact appears in this case.\label{fig:ReducLOCIKLIP}}
   \end{figure}

\subsection{Reduction}\label{Sec:reduc}

Once a reference cube has been assembled and carefully prepared, it can be used in the second main part of our pipeline, to reduce its science images. This part of the pipeline is program-oriented: once reduction parameters have been set by the user, all the images of a chosen program in the given reference cube are reduced altogether with the same parameters.

The ALICE pipeline has been designed to optimally process a large amount of data (the entire NICMOS coronagraphic archive is about 5600 images). This has motivated the choice of the KLIP algorithm\cite{Soummer2012} to reduce the data, 
which is based on principal component analysis and consists only of linear algebra operations. The typical computation time to reduce an image with KLIP is $\sim1$~ms.

This unique celerity enables the exploration of vast parameter spaces to find the optimal reduction for each target in reasonable amount of time. Our pipeline can explore seven different parameters\notspie{ (see Fig.~\ref{fig:ParamSpace})}:
\begin{description}
\item [Mode:] The reference libraries can be optimized depending on the type of structure the user wants to detect. If a disk is being searched in a specific image, all the other images of this target have to be excluded from the reference library to avoid self-subtraction of the potential disk (especially for face-on disks). If planets are being searched, images of this target with sufficently different telescope orientation to avoid self-subtraction can be used in the reference library. 
This is useful to include  PSF images very similar to the target image being reduced. The ALICE pipeline supports both possible modes ("disk" and "planet") and optimizes the PSF library for each individual image being reduced.

\item [Zone:] The reduction process can be applied on specific zones of the science images instead of the full ALICE field of view. 
In the ALICE pipeline, portions of annuli centered on the star can be defined with four parameters (inner radius $R_{min}$, outer radius $R_{max}$, position angle of the center of the zone $CenterPA$, and zone azimuthal length $\Delta PA$). \notspie{Zones position angles (PA) are defined to the North, and rotate with the telescope orientation. This way, should a companion be in the zone field of view, it would be visible in every image of this target whatever the telescope orientation.}

\item [Target preselection:] The reference libraries can be even more optimized to the individual images being reduced:  with this parameter, we can change the size of the PSF library by selecting the reference images the most correlated with the science image.
\notspie{Depending on this value, the effective PSF library can thus be smaller, but more appropriate to reduce a specific image.} This selection might for instance automatically exclude stars with spectral types too different, and increase robustness to any other difference in the PSF (focus change, poor alignment etc.)
\notspie{We compute the correlation between images with a simple cost function computing the quadratic distance between the two images. }
\item [Truncation number:] The last parameter defines the number of KL modes used to create the synthetic PSF subtracted to the science images. A parameter space with truncation numbers spanning the whole range of possible value is particularly useful to optimize the signal-to-noise ratio (SNR) on the reduced image and its possible circumstellar material, and maximize the contrast limits.
\end{description}
A GUI can be used to instantaneously visualize an image of the cube reduced with varying parameters to get a sense of the parameter space to be used for all the targets of the program.

Once a parameter space has been defined by the user, all the images of the selected program are reduced. The computation is parallelized by distributing the reduction of the different targets to the kernels of the machine and accelerate the process. 
\notspie{Routine parameter spaces are usually composed of both "planet" and "disk" modes, one big zone with the entire ALICE field of view except for a central mask of $\sim6$~pixel radius, a handful of criterion values between 0.1 and 0.9, and all the truncation numbers offered by the reference cube length.} For a medium size reference cube ($\sim 450$~images), each science image is reduced with typically $\sim 2250$ different parameter sets, in about 2.8~s. A three-dimension FITS cube is exported for each science image, with the reduced images for all the parameter sets, and can thus be analyzed later or by a different user. \notspie{The cube export is the most time-consuming part of the process, with typically 20~s to export a 2250-image cube on remote STSCI servers.So time 450 images.}

\subsection{Analysis}

\begin{figure}
   \begin{center}
   \begin{tabular}{c}
   \includegraphics[width=21cm,angle=-90]{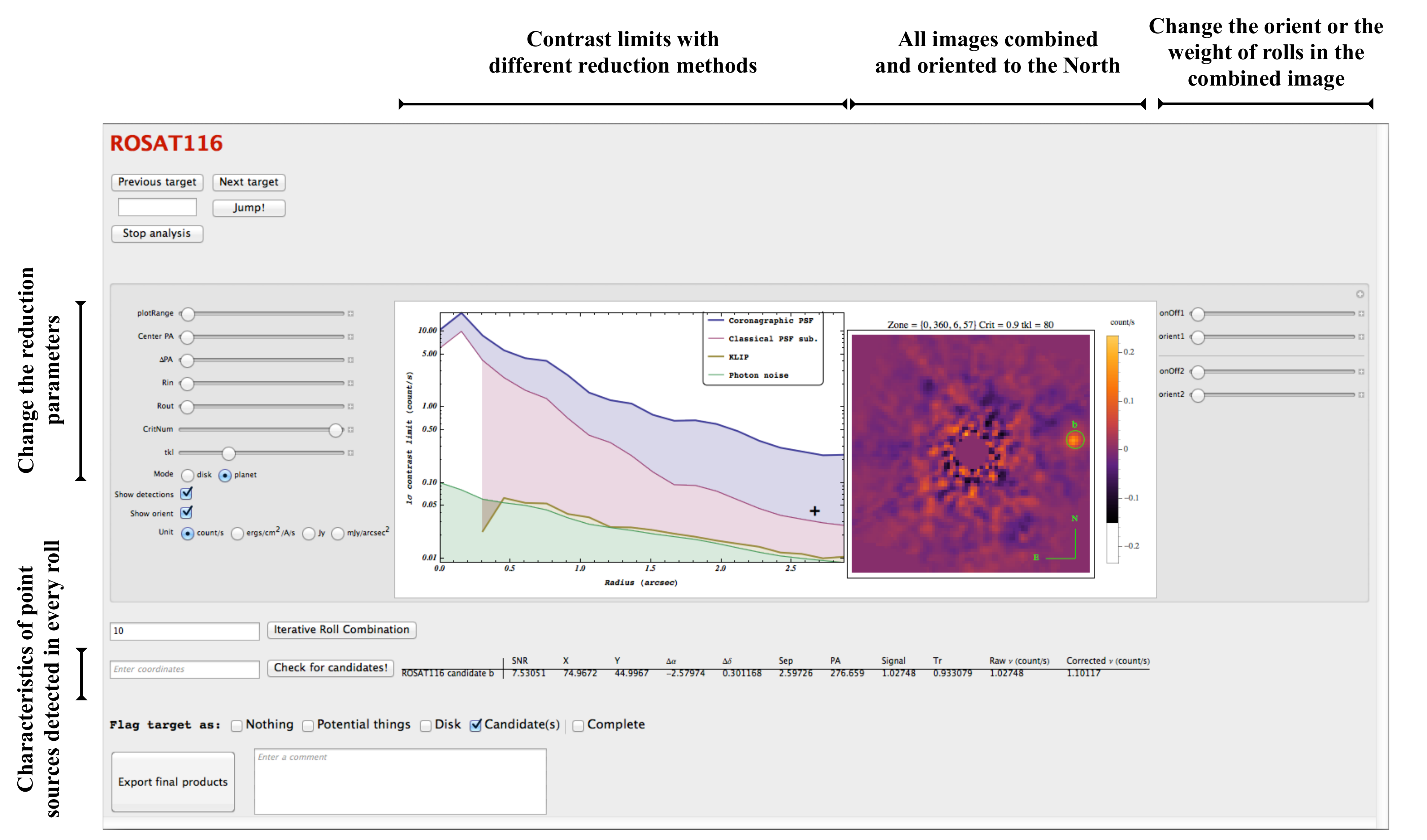}
   \end{tabular}
   \end{center}
   \caption{GUI used to for the analysis of a target (TWA 6 - ROSAT116 in this example). \label{fig:CharliGUI}}
   \end{figure}

The final section of the ALICE pipeline is the analysis part. In this section, reduced image cubes of selected targets are re-imported and individually observed to find circumstellar material with a dedicated GUI (see Fig.~\ref{fig:CharliGUI}).

For a given target being analyzed, all the images are rotated to have the North up, then are co-added and normalized to the total exposure time on the target. The reduction parameter space is manually explored by the user to find the optimal parameters for the combined image. The GUI proposes several tools to help the user choose the best parameter set:
\begin{itemize}
\item The user can change the orientation of a roll (combination of all images acquired with identical telescope orientation) or its weight in the displayed combined image. This enable to visually verify that a point source or a disk is in every roll, by checking how it rotates or decreases in intensity, respectively, in the combined image.
\item Contrast curves are computed for four different combined images: the target raw combined image, the image processed  classical PSF subtraction (median image of the reference cube), the combined image processed with KLIP\notspie{ with the observed parameter set}, and the photon noise estimated from the raw images. The \notspie{three first} contrast curves are computed from the standard deviation in concentric annuli of width 2~pixels ($\sim\lambda/D$). \notspie{The photon noise is computed from the raw images with the squared root of the mean value in the same concentric rings.  Changing the parameter space will automatically update the contrast curve for the KLIP-reduced image, providing the user with a measure of the optimal parameters.}
\end{itemize}
Each individual reduced images can also be analyzed independently throughout the parameter space before combining them in the final image. If the some are not properly reduced (e.g. remaining bad pixel, misalignment), the user can \notspie{flag them and} exclude them from the combined image.

Once the optimal parameter set has been determined, the user can analyze and characterize residual point sources. If the target has been observed in several telescope orientations, detecting a companion at the same sky-coordinates in several rolls not only increases its SNR in the combined image but it also significantly decreases the false-alarm probability. 

For each point source spotted by the user, the pipeline computes its precise astrometry and photometry in each roll, from the cross-correlation with a model PSF. Point sources with failed cross-correlation in at least one roll are considered as residual speckles and are discarded from the candidate list. \notspie{The failed cross-correlation criteria are either a negative photometry or position farther than $\lambda/D$ away from the position spotted by the user.} Each candidate is then characterized in each individual reduced image, in each roll, and in the combined image, by computing:
\begin{itemize}
\item Its astrometry (both in pixel and sky coordinates relative to the star center).
\notspie{in pixel coordinates, and in sky coordinates relative to the star both in right ascension/declination and in separation/position angle. The sky coordinates are corrected from the fact that NICMOS NIC-2 pixels are slightly rectangular.} 
\item Its photometry, both in the reduced images and corrected from the algorithm throughput. The throughput is computed with forward modeling of the candidate, by computing the flux of a model PSF after subtraction by its projection onto the same KL vectors used for the reduced science image (see [\citenum{Soummer2012}] for the analytic expression of forward modeling with the KLIP algorithm). 
\notspie{The algorithm throughput is rigorously identical for images with the same telescope orientation, since the PSF library does not change (all the images of the science target with the same telescope orientation than the image being reduced are excluded from the PSF library, see Sec.~\ref{Sec:reduc}) and consequently the KL vectors do not change either.}
\item Its SNR, by estimating the noise with the standard deviation in a small aperture of radius $3~\lambda/D$ around the candidate and masking it out with an obstruction of radius $\lambda/D$.
\end{itemize}

\begin{figure}
   \begin{center}
   \begin{tabular}{c}
   \includegraphics[width=0.9\linewidth]{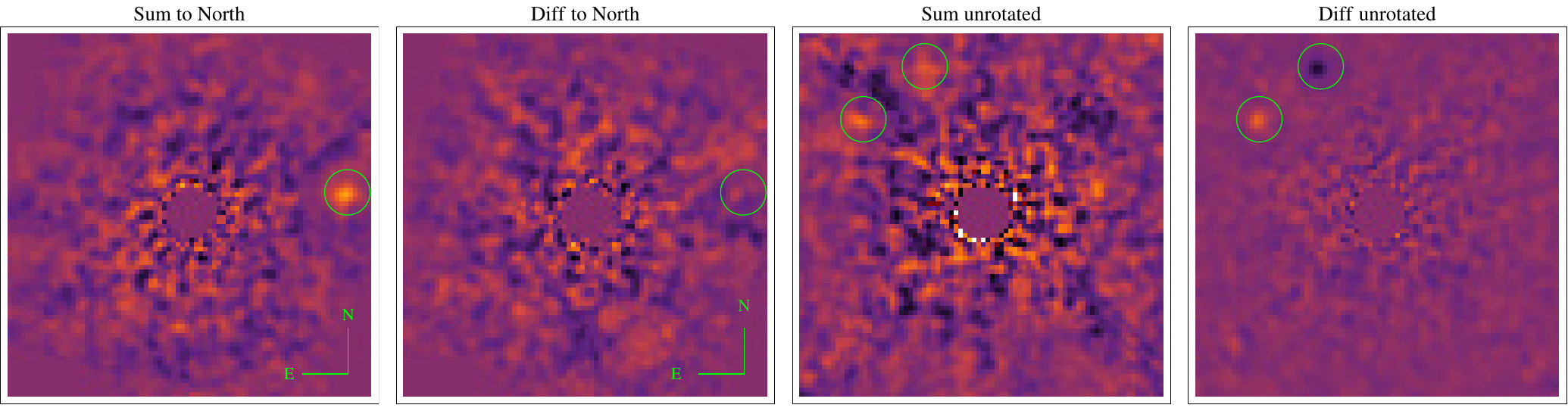}
   \end{tabular}
   \end{center}
   \caption{Combination of binned rolls exported in the FITS cube for the target TWA~6. \textbf{Left:} sum of the binned rolls oriented to the North; the candidate appears with a high SNR. \textbf{Center-left:} Difference between both binned rolls oriented to the North; the candidate auto-subtracts and disappears from the image. \textbf{Center-right:} Sum of the binned rolls in their original orientation; the candidate appears at two different position angles. \textbf{Right:} difference of the two binned rolls in their original orientations; the candidate appears in positive and negative at two different locations.\label{fig:cubeExport}}
   \end{figure}

Once the user is satisfied with the final image and the point source detections, high-level science products are exported, ready to be released to the MAST archive:
\begin{itemize}
\item Each individual reduced image in as many FITS files, with keywords detailing the pipeline parameters.
\item The KL vectors and the zone mask for each image.
\item A FITS cube with $4+N$ images with $N$ the number of rolls. The $N$ \notspie{last} images are the individual rolls, combining all the reduced images acquired with the same telescope orientation. \notspie{The four first other images are tools to visually identify circumstellar material (disk or point sources), computed with different combination of rolls. If the target has been observed with more than two telescope orientations, the reduced roll images are separated in two halves within which the rolls are rotated to the same orientation and combined, giving thus two binned rolls at two different orientations.} The four first images of the exported cube correspond to 1 and 2: the sum and difference of the two rolls both\footnote{If the target has been observed with more than two telescope orientations, the roll images are rotated and combined to provide two binned rolls.} rotated to the North respectively, 3 and 4: the sum and difference of these bins with their original orientations  (See Fig.~\ref{fig:cubeExport}). If point sources or edge-on disks are present in the field of view, they appear at high SNR in the first image and self-subtract in the second, and appear twice with different orientations in the third and fourth images, respectively in positive-positive and positive-negative.
\item A report with the reduced images and the candidates characteristics in the combined image, each individual roll, and each individual image.
\end{itemize}
A standard format is currently being developed for direct imaging science product to enable consistent data interchange between different instruments (Choquet et al., these proceedings (9143-199)), in a specific FITS format including the final reduced image, contrast and SNR maps, and specific FITS keywords. The ALICE pipeline is currently being adapted to export high level science products in this format, that will be then exported to the MAST archive.

\section{FIRST RESULTS}\label{sec:results} 

Although the pipeline has been constantly in development since the beginning of the ALICE project in 2012, we were still able to use it to reduce and partially analyze most of the NICMOS coronagraphic archive. All the targets observed with the F110W and F160W filters (about 70~\% of the database) have been aligned in reference cubes, and at least preliminarily reduced to find interesting targets and targets with obvious circumstellar material. As soon as the pipeline is finalized with all features described herein, the high-level science products will be generated and delivered to the MAST archive.

\subsection{Resolved Structures}

\begin{figure}
   \begin{center}
   \begin{tabular}{c}
   \includegraphics[width=0.95\linewidth]{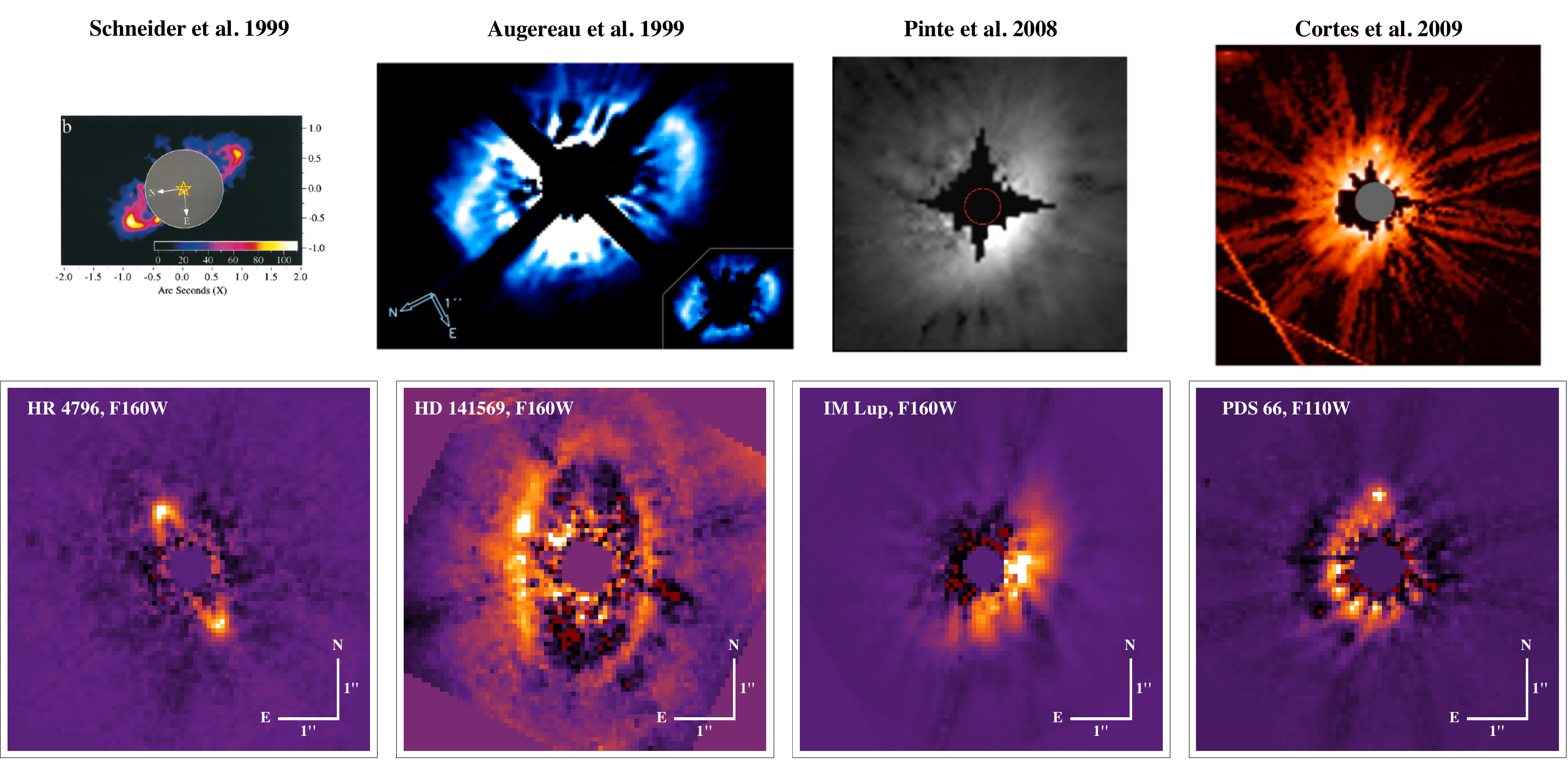}
   \end{tabular}
   \end{center}
   \caption{Known debris and protoplanetary disks imaged in scattered light with the coronagraphic mode of NICMOS. Top row: discovery images. Bottom row: reprocessing with the ALICE pipeline, with the exact same data-set. From left to right: HR 4796A (program 7233, F160W, discovery image from [\citenum{Schneider1999}]) , HD 141569 (program 7857, F160W, discovery image from [\citenum{Augereau1999}]), IM Lup (program 10177, F160W, discovery image from [\citenum{Pinte2008}]), PDS 66 (program 10527, F110W, discovery image from  [\citenum{Cortes2009}]).\label{fig:knownDisks}}
   \end{figure}

The most unanticipated part of the ALICE project is its ability to reveal extended structures around a great number of targets. This can be attributed to two main reasons:
\begin{itemize}
\item The KLIP algorithm, which has good robustness to extended structures and is fast enough to explore very large parameter spaces in a few seconds, and find optimal parameters to reveal circumstellar material.
\item The use of very large PSF libraries, which enables us to be very selective on the similarity between the references and the science image. When working with a 300-image library, discarding the 60~\% less correlated with the science image still provides us with a cube of 120 reference PSF. In addition, since all of  these are very good matches to the science image PSF, the algorithm is very efficient at removing PSF residuals at very small angular separations from the star, down to the photon noise level.
\end{itemize}
With the ALICE pipeline, we were able to easily re-image known disks and improve their detection especially at small separations. Fig.~\ref{fig:knownDisks} present a few of these new reductions for four known disks. These new reductions enabled further analysis and characterization for some of these targets (Milli et al., submitted).

\begin{figure}
   \begin{center}
   \begin{tabular}{c}
   \includegraphics[width=0.95\linewidth]{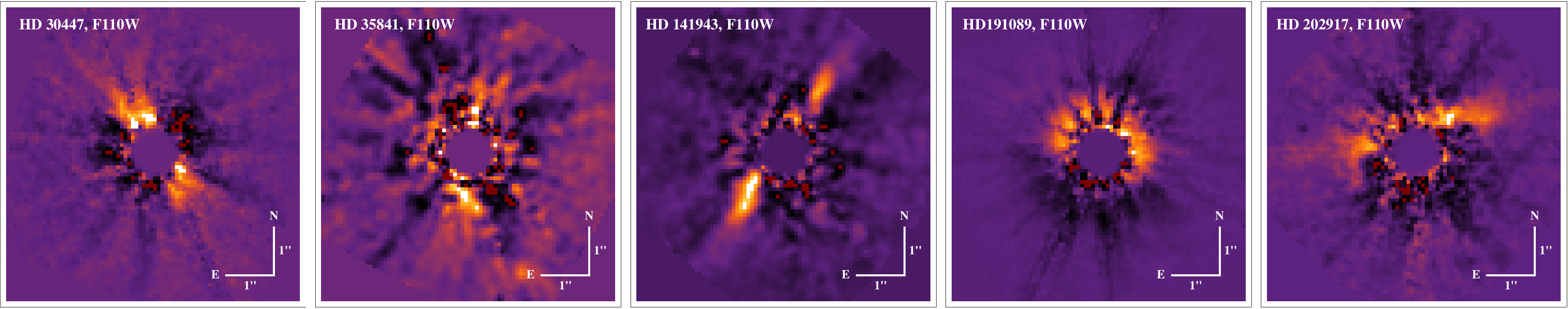}
   \end{tabular}
   \end{center}
   \caption{Five debris disks newly revealed in scattered light  from NICMOS coronagraphic archive (filter F110W) with the ALICE pipeline\cite{Soummer2014}. From left to right: HD~30447 (F3V, 80~pc, 10-40~Myr in Columba association), HD~35841 (F3V, 96~pc, 10-40~Myr in Columba association), HD~141943 (G2V, 67~pc, 17-32~Myr), HD~191089 (F5V, 52~pc, 8-20~Myr in $\beta$ Pictoris association), HD~202917 (G7V, 43~pc, 10-40~Myr in Tuc-Hor association).\label{fig:5disks}}
   \end{figure}

In addition to these previously known disks that are easily detected because of their brightness and/or angular size, we also revealed five disks in scattered-light for the first time, which were previously unseen in the NICMOS archive\cite{Soummer2014}. Only one of them (HD~202917) had already been marginally detected at low SNR in the visible with the HST Advanced Camera for Surveys (ACS). These discoveries increase by 21~\% the number of debris disks imaged in scattered light. Three of these disks appear edge-on (HD~30447, HD~35841 and HD~141943) and the other two are inclined (HD~191089 and HD~202917) (see Fig.~\ref{fig:5disks}). All five host stars are young (8 to 40~Myr), nearby (40 to 100~pc) main sequence stars (F and G type). In particular, HD~141943 is a close analog to the Sun at the age of terrestrial planet formation.

\subsection{Point Sources}

\begin{figure}
   \begin{center}
   \begin{tabular}{c}
   \includegraphics[width=0.8\linewidth]{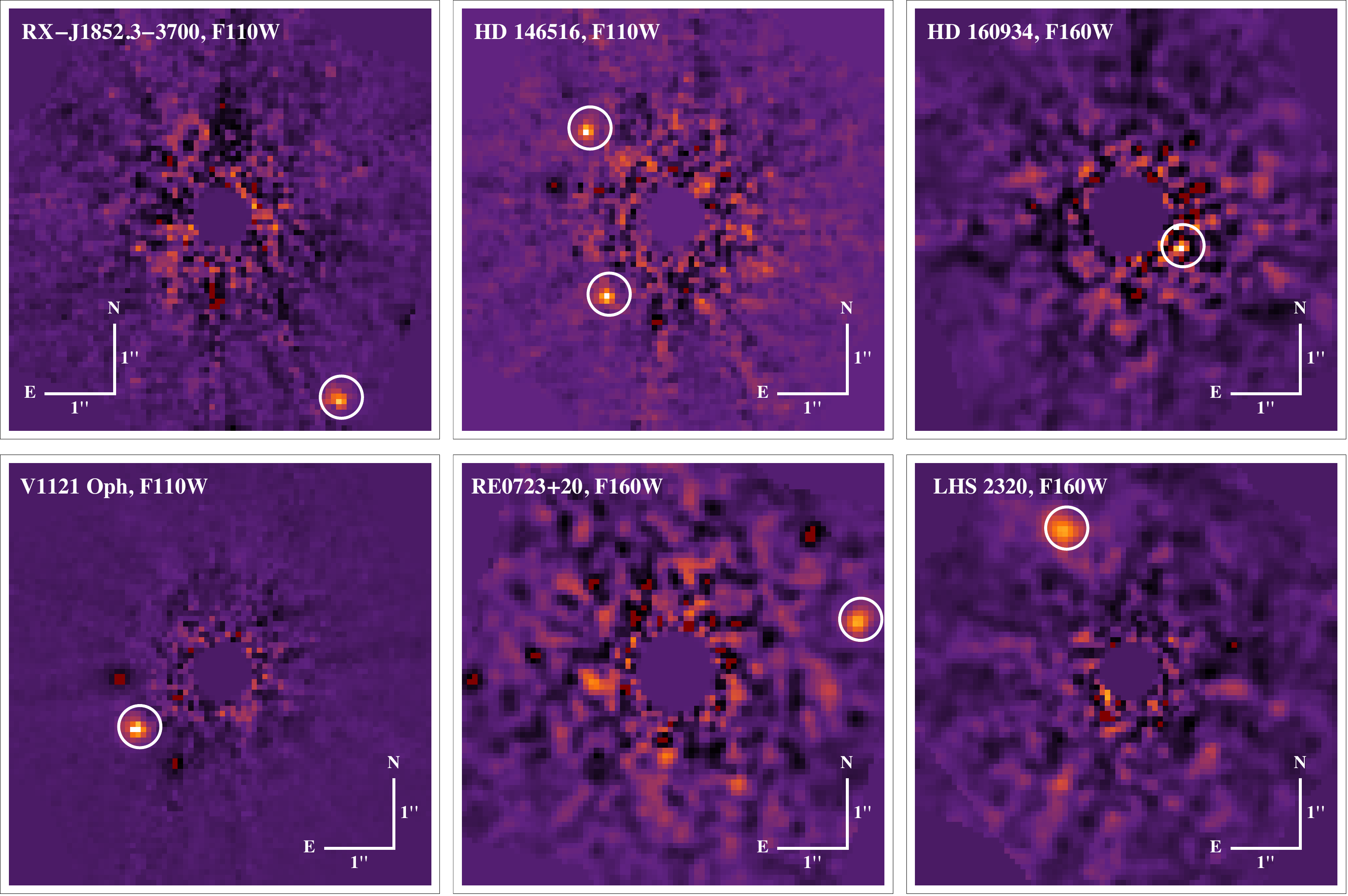}
   \end{tabular}
   \end{center}
   \caption{Six companion candidates discovered in the NICMOS coronagraphic archive with the ALICE pipeline and currently followed-up with the HST-WFC3 instrument (PI: L. Pueyo). From top-left to bottom-right: RX-J1852.3-3700 (K3, 130~pc, 3~Myr in Cha association), HD~146516 (G7V, 120~pc, 10~Myr, in Up. Sco. association), HD~160934 (K7Ve, 33~pc, 100~Myr in AB Dor. association), V1121 Oph (K4Ve, 125~pc, 1~Myr in $\rho$ Oph association), RE0723+20 (K5Ve, 25~pc, 100~Myr in AB Dor. association), LHS~2320 (M5.0V, 22~pc, 100~Myr). \label{fig:candidatesWFC3}}
   \end{figure}

Additionally to these disks, we also imaged and characterized a large number of point sources. A number of them are very close binary systems with high magnitude difference, and were not previously detectable with classical techniques. We also found a dozen of substellar companion candidates, which have yet to be confirmed as co-moving with the host star using followup observations. Fig.~\ref{fig:candidatesWFC3} presents six candidates which are currently followed-up with HST-WFC3 to confirm their gravitational bound with the host star (PI: L. Pueyo). From stellar populations models, we estimated a probability of 98~\% to have at least one gravitationally-bound companion among this sample. Followup observations of the other point source candidates that are bright enough to be observed with ground-based AO systems are also currently being conducted to confirm companionship with the host stars.

\section{CONCLUSION} \label{sec:conc}

In this paper, we presented the ALICE project, which consists in a comprehensive and consistent reprocessing of the HST-NICMOS coronagraphic archive using advanced PSF subtraction algorithms. We briefly described the content of the archive, then provided a detailed description of the ALICE pipeline. Finally, we presented a summary of the first scientific results of this project obtained with a preliminary version of the pipeline, showing new images of debris disks in scattered light as well as unknown potential sub-stellar companions.

The pipeline is currently in its final development phase, consisting mostly in the optimization of the image alignment process and in the definition and implementation of a standard format for the high-level science products that will be generated for all the images in the archive. The entire archive will then be consistently reprocessed with the final version of the pipeline and the final products will be delivered to the MAST archive.

All of our discoveries are now included in followup observation campaigns. Follow-up observations of the point source candidates are being conducted with the HST-WFC3 for the six faintest targets, and with ground-based instruments (VLT-SINFONI, Keck, Palomar P1640) for the brigher ones others, to confirm their companionship and characterize their photometry and colors. In addition, an HST-STIS coronagraphic program is currently on-going to observe the five debris disks newly imaged with the ALICE pipeline, and will provide visible images of the disks by Fall 2014, with a resolution twice better than with NICMOS (PI: M.~D. Perrin). These five disks are currently being analyzed by our team using the radiative transfer modeling MCFOST\cite{Pinte2006} to characterize their optical properties and composition.

\acknowledgments 
This project was made possible by the Mikulski Archive for Space Telescopes (MAST) at STScI. Support was provided by NASA through grants HST-AR-12652.01 (PI: R. Soummer), HST-GO-11136.09-A (PI: D. Golimowski), and by STScI Director’s Discretionary Research funds, from STScI, which is operated by AURA under NASA contract NAS5- 26555. The input images to ALICE processing are from the recalibrated NICMOS data products produced by the Legacy Archive project, “A Legacy Archive PSF Library And Circumstellar Environments (LAPLACE) Investigation,” (HST-AR- 11279, PI: G. Schneider). 
Pueyo was supported in part under contract with the California Institute of Technology (Caltech) funded by NASA through the Sagan Fellowship Program executed by the NASA Exoplanet Science Institute.
This research has made use of the SIMBAD database, operated at CDS, Strasbourg, France.


\bibliography{biblio_disk-planet,biblio_haut-contraste}   
\bibliographystyle{spiebib}   

\end{document}